# Numerical simulations of Josephson Traveling Wave Parametric Amplifiers (JTWPAs): comparative study of open-source tools

A. Yu. Levochkina, H. G. Ahmad, P. Mastrovito, I. Chatterjee, D. Massarotti, D. Montemurro, F. Tafuri, G.P. Pepe and M. Esposito

*Abstract*— Josephson Traveling Wave Parametric Amplifiers (JTWPAs) are largely exploited in quantum technologies for their broadband and low noise performance in the microwave regime. When one or more microwave tones are applied at the input, such devices show a complex wave-mixing response due to their intrinsic nonlinear nature. Numerical simulations of the JTWPAs nonlinear behaviour provide useful insights not only for the design of such devices, but also for the interpretation and validation of the experimental results. Here we present and discuss a comparative analysis of different open-source tools which can be used for JTWPAs numerical simulations. We focus on two tools for transient simulations, WRSPICE and PSCAN2, and on one tool for direct simulation of the frequency domain behaviour, JosephsonCircuit.jl. We describe the working principle of these three tools and test them considering as a benchmark a JTWPA based on SNAILs (Superconducting Nonlinear Asymmetric Inductive eLement) with realistic experimental parameters. Our results can serve as a guide for numerical simulations of JTWPAs with open-source tools, highlighting advantages and disadvantages depending on the simulation tasks.

## I. INTRODUCTION

Quantum-limited microwave parametric amplifiers are widely used in quantum technologies applications, ranging from signal redout improvement, e.g. in context of superconducting qubits, to generation of microwave squeezing [1,2]. Josephson Parametric Amplifiers (JPAs) based on Josephson junctions built into a resonant circuit provide quantum-limited amplification but at the same time their resonant nature limits their bandwidth [3]. Josephson Traveling-Wave Parametric Amplifiers (JTWPAs) allow to overcome the bandwidth restrictions due to their non-resonant design [4-19]. A JTWPA is a non-linear transmission line composed of an array of Josephson junctions-based repeating elements (unit cells). Depending on the kind of repeating element that is used as unit cell, one can identify different types of such amplifiers, e.g., single Josephson junction-based TWPAs [4, 9, 10], SQUID (Superconducting Quantum Interference Device) [7, 8, 11, 12, 13, 14, 15] or asymmetric SQUID-based TWPAs [5, 16, 17, 18, 19].

Amplification in JTWPAs is typically described by means of analytic models based on coupled mode equations (CMEs). The CMEs approach provides a simple and quick way to understand the key feature of the gain mechanism in JTWPAs, but it does not fully capture the broadband nature of these devices since it considers only a limited number of propagating tones. Numerical circuit simulations can instead provide a more realistic description of the frequency dependent response of JTWPAs and recently started to be extensively adopted for JTWPAs research [10, 11, 12, 13, 14].

In this work we report on a comparative study of open-source numerical tools for simulating JTWPAs.

Specifically, we provide accurate comparison between different open-source numerical tools in terms of operating principles, circuits parametrization, simulation speed and methods for implementing flux tunability. As a benchmark for our analysis, we consider a JTWPA based on SNAIL [20] unit cells with realistic experimental parameters [16].

## II. BENCHMARK DEVICE

We now briefly describe the benchmark device adopted for our study. In Fig. 1 a sketch of the considered device is shown,

This work is supported by the European Union under Horizon Europe 2021-2027 Framework Programme Grant Agreement no. 101080152 and under Horizon 2020 Research and Innovation Programme Grant Agreement no. 731473 and 101017733; and by PNRR MUR project PE0000023-NQSTI.

A.Yu. Levochkina Department of Physics University Federico II, Naples, 80126 Italy and CNR-SPIN Complesso di Monte S. Angelo, via Cintia, Napoli 80126 (e-mail: anna.levochkina@unina.it)

H.G. Ahmad Department of Physics University Federico II, Naples, 80126 Italy and CNR-SPIN Complesso di Monte S. Angelo, via Cintia, Napoli 80126 (e-mail: halimagiovanna.ahmad@unina.it)

P. Mastrovito Department of Physics University Federico II, Naples, 80126 Italy and CNR-SPIN Complesso di Monte S. Angelo, via Cintia, Napoli 80126 (e-mail: pasquale.mastrovito@unina.it)

I. Chatterjee Department of Physics University Federico II, Naples, 80126 Italy and CNR-SPIN Complesso di Monte S. Angelo, via Cintia, Napoli 80126 (e-mail: isita.chatterjee@unina.it)

D. Massarotti Department of Electrical Engineering and Information Technology University Federico II, Naples, 80126 Italy and CNR-SPIN Complesso di Monte S. Angelo, via Cintia, Napoli 80126 (e-mail: davide.massarotti@unina.it)

D. Montemurro Department of Physics University Federico II, Naples, 80126 Italy and CNR-SPIN Complesso di Monte S. Angelo, via Cintia, Napoli 80126 (e-mail: domenico.montemurro@unina.it)

F. Tafuri Department of Physics University Federico II, Naples, 80126 Italy (e-mail: francesco.tafuri@unina.it)

G.P. Pepe Department of Physics University Federico II, Naples, 80126 Italy (e-mail: pepeg@na.infn.it)

M.Esposito CNR-SPIN Complesso di Monte S. Angelo, via Cintia, Napoli 80126, Italy (e-mail: martina.esposito@spin.cnr.it)

Software codes used in this work are available [Online]: A. Levochkina, 2023, http://github.com/levochkinanna/JTWPA_Numerical_Simulations

where the SNAIL unit cell is illustrated in the inset. A SNAIL is an asymmetric SQUID composed of a superconducting loop with $N$ large junctions in one arm and one small junction in the other arm.

In the considered device, SNAILs are oriented in such a way that the external magnetic flux has opposite sign for adjacent SNAIL cells. The aim of such alternating flux polarity design is to mitigate the overall three-wave-mixing nonlinearity, which is an odd function of the external flux. This kind of device is typically operated in the regime of four-wave-mixing amplification [16, 17]. In this regime, in presence of an intense input pump at frequency $f_p$ and a weak input signal at frequency $f_s$, the output signal will be amplified and an idler tone at frequency $f_i = 2f_p - f_s$ will be generated.

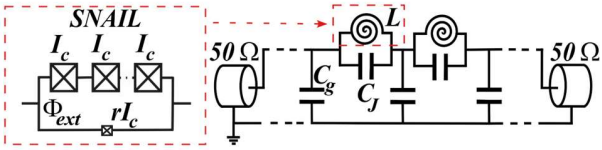

**Fig. 1.** Sketch of a SNAIL TWPA. $I_c$ is the critical current of the large Josephson junction; $r$ is the ratio between small and large Josephson junctions, $L$ is the inductance per unit cell, $C_J$ is the Josephson capacitance and $C_g$ is the ground capacitance. The SNAIL symbol is sketched with opposite orientation for adjacent cells, indicating the alternating flux polarity design.

Our benchmark device consists of 250 SNAILs. Experimentally realistic parameters for the SNAIL TWPA are considered [16] and reported in Table 1.

TABLE I
REALISTIC PARAMETERS FOR A SNAIL TWPA DEVICE [16]

| Parameters | Value |
|---|---|
| number of cells | 250 |
| critical current ($I_c$) | 1.47 $\mu A$ |
| ratio between small and large JJs ($r$) | 0.05 |
| Josephson capacitance ($C_J$) | 31 $fF$ |
| ground capacitance ($C_g$) | 550 $fF$ |

Following the CMEs theoretical approach described in [17], the four-wave-mixing gain for such device can be analytically computed. As a reference, in Fig. 2 we report the CMEs computed gain as a function of pump power and signal frequency without including losses in the model. The pump frequency $f_p$ was set at 4.415 GHz and two significative values of the external flux were considered, zero and half flux quantum $\Phi_0$.

In the following sections we describe in detail how to implement JTWPA numerical simulations with three open-source tools: PSCAN2 [21], WRSPICE [22] and JosephsonCircuit.jl [23]. We test the three tools by using them to model four-wave mixing gain for the benchmark SNAIL TWPA device described above. Specifically, we simulate four-wave-mixing gain for two significative values of the external flux, zero and half flux quantum $\Phi_0$, and we compare simulation methodologies and performance.

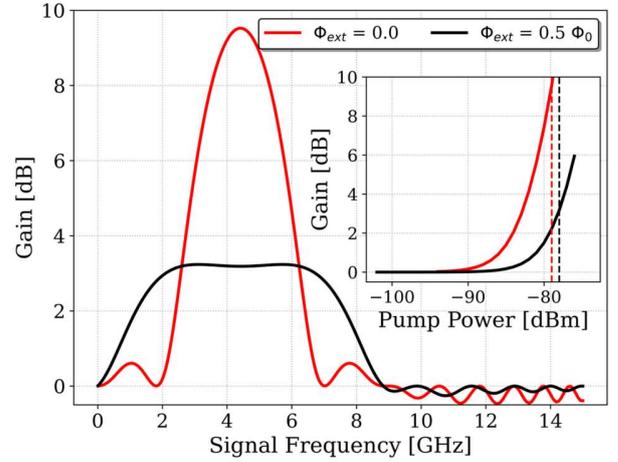

**Fig. 2.** Analytically computed four-wave-mixing gain using CMEs theory for a SNAIL TWPA with parameters described in the Table1. For 0.0 external flux, pump power is set to -79 dBm (*red*) and for 0.5 $\Phi_0$ external flux, pump power is set to -78dBm (*black*). The inset shows gain as a function of pump power when the signal frequency is $f_s$ = 4.215 GHz. Pump frequency is $f_p$ = 4.415 GHz.

III. NUMERICAL SIMULATIONS

There are several circuit simulation tools allowing to simulate Josephson junctions-based devices, e.g., spice-based simulators such as WRSPICE [22], JSIM [24] and JoSIM [25]; PSCAN [26] and PSCAN2 [21]; and JosephsonCircuit.jl [23] implemented for the *Julia* programming language. Here we perform a comparative analysis of three different open-access simulators, WRSPICE, PSCAN2 and JosephsonCircuit.jl, that recently started to be extensively adopted for simulating superconducting quantum devices [10, 11, 14, 27, 28]. For each of them we describe the key operating principles and discuss the performance to provide a guidance for identifying the optimal tool according to the user's simulation needs. All simulations are performed using a laptop PC11[th] Gen Intel® Core ™ i3-1125G4 with 8 GB RAM and 4-core processor with bus speed 4GT/s.

*A. Transient Simulations: basic concepts*

PSCAN2 and WRSPICE simulators can be used to perform so-called transient circuit simulations (the analysis of the circuit electromagnetic response as a function of time for a given input). The basic concept of transient circuit simulations is schematically illustrated in Fig. 3.

For simulating the gain of a TWPA, the input electromagnetic wave must be composed of two frequency tones: a pump tone with high amplitude (which provides the energy necessary for the amplification) and a signal tone with low amplitude (which is expected to be amplified at the output).

The simulators allow to register the current at each node of the circuit as a function of time. In our case, we are specifically interested in registering the current at the device input and output.

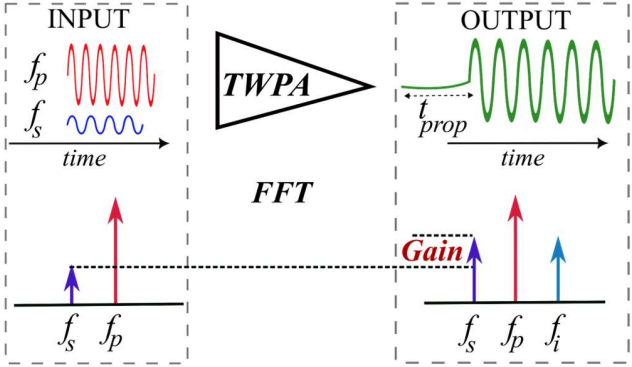

**Fig. 3.** Schematic illustration of the transient circuit simulation basic concept. Here $f_p$, $f_s$, and $f_i$ are frequencies of pump, signal and idler tones, respectively and $t_{prop}$ is the propagation time.

A way to check that this kind of transient circuit simulations are working properly is to verify that the propagation time, namely the time necessary to see a nonzero signal appearing at the output, matches with the expected one, which can be calculated with simple transmission line theory. For the device under test the propagation time is $t_{prop} = N\sqrt{LC_g} = 4.406$ ns (where $N$ is the number of unit cells, $L$ is the inductance per unit cell and $C_g$ is the ground capacitance per unit cell).

In Table 2, we report the experimentally realistic parameters [16] that we adopted for this kind of simulations. The adopted pump powers are chosen to investigate the device in a regime of stable gain. Once, for a given *pump + signal* input, the simulation converges, we perform the Fourier Transform of the output and input signals and from their difference we extract the TWPA gain at the selected signal frequency [14]. We consider the propagation time as the starting time for the data to be processed with the Fourier Transform. The same procedure is then repeated for several signal frequencies to obtain the gain in the desired frequency region. This general methodology is identical for PSCAN2 and WRSPICE. For both tools, we run gain simulations for 51 values of signal frequency in the range from 0 to 10 GHz. In the following we describe each tool in more detail.

TABLE 2
SIMULATION SIGNAL AND PUMP PARAMETERS

|  | 0.0 flux quantum | 0.5 flux quantum |
|---|---|---|
| **pump frequency ($f_p$)** | 4.415 *GHz* | 4.415 *GHz* |
| **signal frequency ($f_s$)** | varying | varying |
| **pump power** | -79 *dBm* | -78 *dBm* |
| **signal power** | -110 *dBm* | -110 *dBm* |

### B. PSCAN2 simulator

PSCAN2 (Portable Superconductor Circuit Analyzer) is a software package that allows to analyze superconducting electronics circuits. The analysis of superconducting circuits is carried out based on the Josephson phase balance equations for the superconducting circuits, i.e., by *the nodal phase method*. The high efficiency of the PSCAN2 simulator is achieved using an automatically varying step of integration (in time) in the numerical solution of system's differential equations. This system of equations is generated within the software package automatically based on the given structure of the circuit under study and the used models of Josephson elements: RSJ (resistively shunted) or TJM (tunnel junction model) [21]. To describe the structure of the circuit, a special internal language is used, which includes symbols for the models of Josephson junctions and a variety of circuit elements such as resistors, capacitors, superconducting inductors, as well as voltage, current and phase generators.

For the SNAIL TWPA simulations, the possibility of including phase generators as circuit elements allows to easily set the external magnetic flux for each unit cell. The equivalent circuit for the SNAIL TWPA simulation in PSCAN2 environment is sketched in Fig. 4.

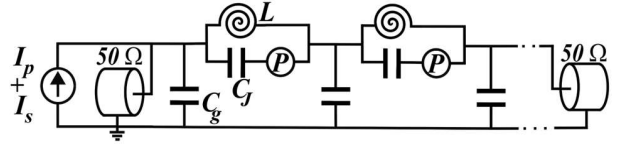

**Fig. 4.** Sketch of the equivalent circuit of a SNAIL TWPA in PSCAN2 environment. *P*-symbols indicate phase generators.

A distinctive feature of this tool, that needs to be carefully considered, is that all the used physical quantities are defined as *dimensionless*. In particular, the unit for magnetic flux is automatically selected in the PSCAN2 system of units in such a way that the value of the magnetic flux quantum, $\Phi_0 = h/2e$, is equal to $2\pi$. In addition, the user must select the PSCAN2 normalization factors for currents and voltages and for the other used quantities accordingly. The user should thus be careful in establishing the correct correspondence between standard SI units and PSCAN2 units. To provide a practical example, in Table 3 we report the chosen correspondence of units for the circuit parameters of the considered TWPA.

Using the equivalent circuit in Fig. 4 and the RSJ model of Josephson junctions, we simulated four-wave-mixing gain for the SNAIL TWPA device under study considering the external flux equal to zero and half flux quantum $\Phi_0$. The results are reported in Fig. 9 and Fig. 10. For RSJ model of Josephson junctions, PSCAN2 requires in input values of Josephson junction critical current, Josephson junction capacitance and the value of junction's normal resistance. We estimate the normal resistance from the critical current according to Ambegaokar–Baratoff relation, $I_c R_N = \pi\Delta/(2e)$ [29], where $\Delta$ is the superconducting gap (aluminium Josephson junctions are considered).

TABLE 3
PARAMETERS OF SNAIL TWPA IN PSCAN2

| Physical quantity | Actual values | PSCAN2 Normalisation factor | PSCAN2 values |
|---|---|---|---|
| Magnetic flux quantum | $\Phi_0$ | $\Phi_n = \dfrac{\Phi_0}{2\pi}$ | $2\pi$ |
| $I_C$ | $1.47\ \mu A$ | $I_n = 1\ \mu A$ | $I_C/I_n$ |
| $I_C R_N$ | $0.33\ mV$ | $V_n = 1\ mV$ | $I_C R_N/V_n$ |
| $R_N$ | $224.5\ \Omega$ | $R_n = 1\ k\Omega$ | $R_N/R_n$ |
| $L$ (unit cell inductance) | $584\ pH$ | $L_n = \dfrac{\Phi_n}{I_n}$ | $L/L_n$ |
| $C_J$ | $31\ fF$ | $C_n = \dfrac{L_n}{R_n^2}$ | $C_J/C_n$ |
| $C_g$ | $550\ fF$ | $C_n = \dfrac{L_n}{R_n^2}$ | $C_g/C_n$ |
| $\omega_p$ | $2\pi \times 4.415\ GHz$ | $\omega_n = 1/\sqrt{L_n C_n}$ | $\omega_p/\omega_n$ |

We set the total simulation time T = 20 ns, while the time step is set to 10 PSCAN2 time units which corresponds to 3.29 ps. Fig 5. shows an example of simulation for signal frequency $f_s$ = 4.215 GHz. For illustration purpose, the dynamics is shown for the first 10 ns. The top graph is the obtained simulated output signal while the bottom graph shows the corresponding Fourier Transform. This simulation takes 33.0 s to run on a laptop PC (PC11[th] Gen Intel® Core ™ i3-1125G4).

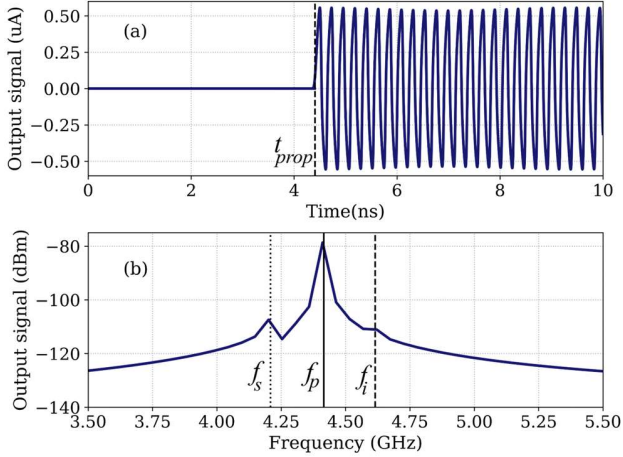

**Fig. 5.** PSCAN2 output time dependent signal (a) and corresponding Fourier Transform (b).

### C. WRSPICE simulator

WRSPICE is a flexible circuit simulator which includes the possibility to model a variety of circuit elements (capacitors, inductors and coupled inductors, resistors, switches, dependent and independent current and voltage sources, Josephson junctions, etc.) [22]. WRSPICE, as well as PSCAN2, offers two Josephson junction models: RSJ and TJM. We used RSJ model of Josephson junction as in PSCAN2. For RSJ model of WRSPICE it is necessary to set the Josephson junction's critical current and its capacitance [30].

In contrast to PSCAN2, WRSPICE does not allow to include phase generator as circuit element. So, to set the desired external flux for each unit cell, we need to modify the SNAIL TWPA WRSPICE equivalent circuit, and model the flux bias by introducing a small linear inductance, $L_{add} = 0.1\ pH$, in each SNAIL element, and adding an auxiliary external inducting loop with mutual inductance $M$ (Fig. 6). In this way, the flux is coupled via the external inductor $L_{ext}$ and is controlled by an external source of dc current $I_{dc}$, which is included in the model. Here we set $L_{ext} = 1.6\ \mu H$ and the mutual inductance is proportional to $\sqrt{L_{add} L_{ext}}$ [19]. The case of 0 external flux corresponds to setting $I_{dc} = 0.0\ \mu A$, while 0.5 external flux corresponds to setting $I_{dc} = -2.5\ \mu A$.

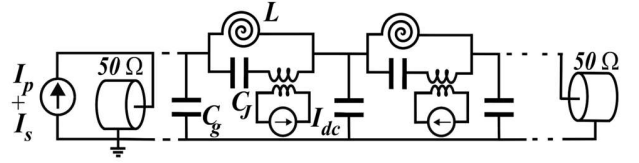

**Fig. 6.** Equivalent circuit of a SNAIL TWPA adopted for WRSPICE environment.

Using the equivalent circuit from the Fig. 6 we simulated four-wave-mixing gain for the SNAIL TWPA device under consideration. The results are reported in Fig. 9 and Fig. 10. The total simulation time (T = 20 ns) and the time step (3.29 ps) are identical to those used for PSCAN2 simulations. The time step value is comparable with values previously adopted in the literature [14]. In Fig. 7 we show an example of simulation results for signal frequency $f_s$ = 4.215 GHz. The simulation takes 40.0 s to run on a laptop PC (PC11[th] Gen Intel® Core ™ i3-1125G4).

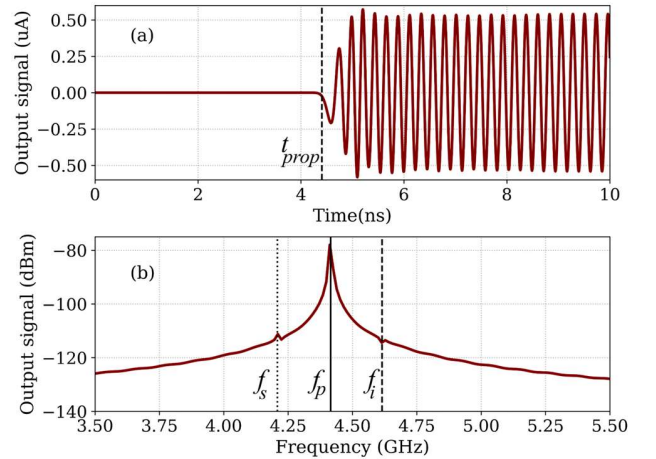

**Fig. 7.** WRSPICE output time dependent signal (a) and corresponding Fourier Transform (b).

## D. JosephsonCircuit.jl

JosephsonCircuit.jl is a package recently developed for the open-source Julia programming language, specifically developed for scientific numerical computing [31]. This package allows high-performance frequency domain simulations of nonlinear circuits containing Josephson junctions [23]. The tool implements fast simulations of scattering parameters and noise in nonlinear circuits going beyond the signal/idler two-mode model and allowing the user to specify the number of modes to account for in the simulation. The frequency domain simulation is based on a variant of nodal analysis [32, 33] and the harmonic balance method [34, 35, 36] with an analytic Jacobian. The Josephson junctions are modeled as lossless nonlinear inductors.

The equivalent circuit that we used for simulations with JosephsonCircuit.jl is the same as the one adopted for WRSPICE (see Fig. 6), with the difference that we do not include auxiliary circuits.

This numerical tool does not currently allow to perform transient simulation, although in the documentation the authors mention that this functionality is under development [23]. Despite the absence of transient simulation capabilities, JosephsonCircuit.jl allows fast calculation of the scattering parameters and quantum efficiency of a specified JTWPA requesting in input pump power amplitude, pump frequency, frequency range and total number of modes to be considered (we set 10 modes). This last feature allows to consider the effect of the interacting sidebands modes, which originates for example from the mixing between the signal and the odd harmonics of the pump in the four-wave mixing regime [10].

In Fig. 8 we report an example of JosephsonCircuit.jl simulation results for the scattering parameters of the considered SNAIL TWPA device.

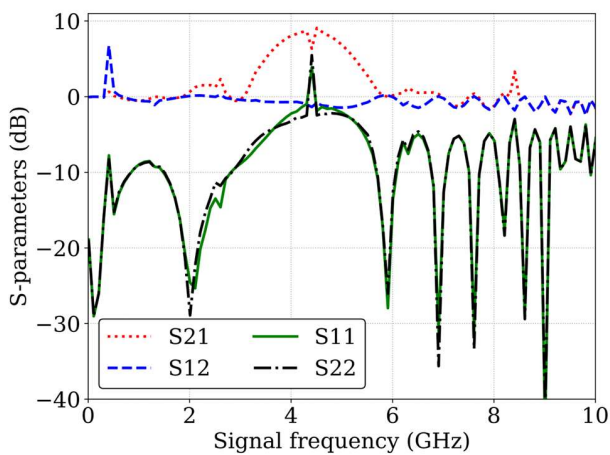

**Fig. 8.** Example of S-parameters as a function of signal frequency for a SNAIL TWPA simulated with JosephsonCircuit.jl. Pump frequency $f_p$=4.415 GHz.

The required input parameters for modelling the Josephson junctions are the critical current and the Josephson capacitance. The gain results obtained with JosephsonCircuit.jl for the case of zero external flux is reported in Fig. 9. We set 106 frequency points in the range 0 to 10 GHz. The full frequency domain simulation takes 17.0 s to run. For this recently developed simulator, we could not consider the case with external flux different from zero, since phase generators (method adopted for PSCAN2) are not available as circuit elements in the package and the possibility to introduce auxiliary inductive circuits (method adopted for WRSPICE) is not currently included in the software documentation.

## IV. DISCUSSION AND CONCLUSIONS

In this work we simulate four-wave-mixing gain in a SNAIL TWPA with realistic experimental parameters [16] by using the three numerical tools described in the previous sections. The results are shown in Fig. 9 and Fig. 10 and compared with standard two-mode CMEs theory.

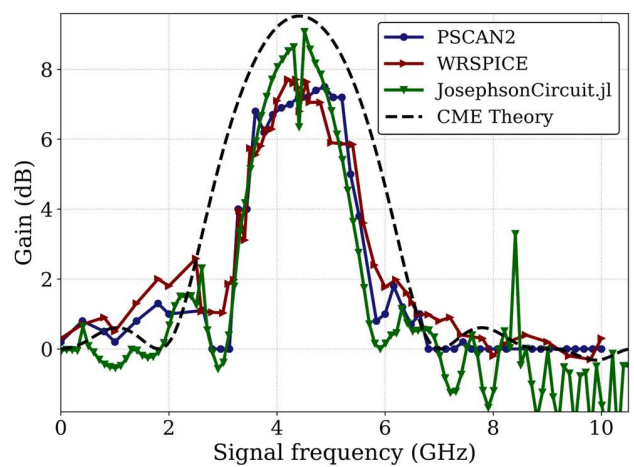

**Fig. 9.** Gain as a function of signal frequency for $\Phi_{ext} = 0$ obtained with different simulation tools and compared with CMEs theory (dashed line). Pump power is -79 dBm, pump frequency $f_p$= 4.415 GHz, signal power is -110 dBm.

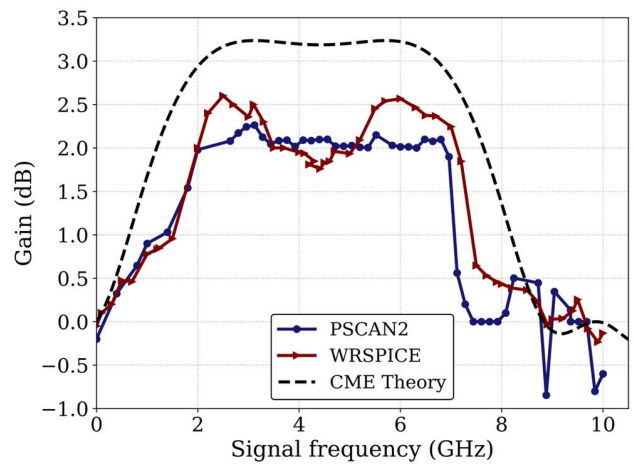

**Fig. 10**. Gain as a function of signal frequency for $\Phi_{ext} = 0.5$ obtained with transient simulation tools and compared with CMEs theory (dashed line). Pump power is -78 dBm, pump frequency $f_p$=4.415 GHz, signal power is -110 dBm.

Both figures show that the results of the numerical simulations are in qualitative agreement among them, while, in comparison to the CMEs theory, the numerical simulations consistently predict a slightly lower gain and narrower bandwidth. This behavior is expected since the CMEs model is a minimal model that does not capture the multi-mode broadband nonlinear behavior of the device.

In terms of tools capabilities and performance, our study highlights the following key aspects.

*PSCAN2* can be used to perform transient circuit analysis with an interface that allows the user to follow the circuit response "live" in time. On the utilized laptop PC, it takes 33.0 s to perform one simulation with total time T=20 ns and time step equal to 3.29 ps. The tool has a normalized system of units, which consider magnetic flux quantum as $2\pi$ that the user should carefully account for. A quite useful feature is the possibility of including phase generators as circuit elements allowing an easy setting of the external magnetic field, particularly relevant for SNAIL TWPAs.

*WRSPICE* simulator also can be used to perform transient circuit analysis of JTWPAs. On the utilized laptop PC, it takes 40.0 s to perform the simulation of a single frequency point for the same parameters used for PSCAN2. To flux bias JTWPAs, this simulator allows the user to include auxiliary circuits, while there are no constraints regarding the choice of units.

Finally, *JosephsonCircuit.jl* is the Julia language package that performs fast frequency-domain simulations of scattering parameters considering an arbitrary number of modes. Our simulation of four-wave-mixing gain in a SNAIL TWPA [16] takes 17.0 s considering 10 modes and 106 frequency points in the selected range.

For the only purpose of investigating the frequency dependent gain of a JTWPA in absence of external magnetic flux, JosephsonCircuit.jl is certainly a much faster tool compared with PSCAN2 and WRSPICE. We stress that the running time of 33.0 s and 40.0 s respectively for PSCAN2 and WRSPICE is referred to the simulation of a single frequency value; the simulation of the full frequency range, considering the same number of frequency points selected for JosephsonCircuit.jl, would take 58 minutes and 70 minutes, respectively for PSCAN2 and WRSPICE.

Transient circuit simulations, however, can provide more flexibility since the input signals can be customized by the user in terms of frequencies and powers combinations, allowing a larger variety of investigations.

Finally, we stress that, although this study focused on simulations of gain only, a very interesting and distinctive feature of JosephsonCircuit.jl package is the possibility to evaluate the quantum efficiency (output to input signal-to-noise-ratio) including an arbitrary number of modes.

In conclusion, we presented a comparative study of open-source numerical tools for the simulation of JTWPA devices with realistic experimental parameters, using as a benchmark a SNAIL-based TWPA device. Our results provide insights into practical aspects, performance, and capabilities of the investigated simulators, facilitating the choice and the usage of the optimal tool depending on the applications.